\documentclass[Afour,sageh,times]{sagej}

\usepackage{moreverb,url}

\usepackage[colorlinks,bookmarksopen,bookmarksnumbered,citecolor=red,urlcolor=red]{hyperref}

\usepackage[flushleft]{threeparttable}

\newcommand\BibTeX{{\rmfamily B\kern-.05em \textsc{i\kern-.025em b}\kern-.08em
T\kern-.1667em\lower.7ex\hbox{E}\kern-.125emX}}

\begin{document}

\runninghead{biniLasso: automated cut-point detector}

\title{biniLasso: Automated cut-point detection via sparse cumulative binarization}

\author{Abdollah Safari\affilnum{1}, Hamed Helisaz\affilnum{2,3}, Peter Loewen\affilnum{4}}

\affiliation{\affilnum{1}School of Mathematics, Statistics, and Computer Science, College of Applied Sciences, University of Tehran, Iran\\
\affilnum{2}Department of Statistics, Faculty of Science, University of British Columbia, Vancouver, Canada\\
\affilnum{3}GranTAZ Consulting LTD, Vancouver, Canada\\
\affilnum{4}Faculty of Pharmaceutical Sciences, University of British Columbia, Canada}

\corrauth{Abdollah Safari,
School of Mathematics, Statistics, and Computer Science,
College of Applied Sciences,
University of Tehran,
Tehran, Iran.}

\email{a.safari@ut.ac.ir}

\begin{abstract}
We present biniLasso and its sparse variant, miniLasso, novel methods for prognostic analysis of high-dimensional data that enable the detection of multiple cut-points per feature. Our approach extends commonly used survival models, with a focus on the Cox proportional hazards models, as well as generalized linear models by integrating a cumulative binarization scheme with $\ell_1$ penalization. The miniLasso variant incorporates an additional uniLasso regularization stage to deliver a more parsimonious model. Both methods are computationally efficient (2–8× faster than existing approaches, on average) and demonstrate improved performance in extensive simulations and applications to three genomic cancer datasets from TCGA. The standard biniLasso excels at uncovering complex relationships in exploratory analyses where the number of cut-points is unrestricted. Crucially, when the model is constrained to a limited number of cut-points, a common requirement for clinical utility, miniLasso achieves comparable predictive accuracy while delivering a simpler, more interpretable model.
\end{abstract}


\maketitle

\section{Introduction}
\label{s:intro}

The discretization of continuous predictors is a cornerstone of interpretable modeling in clinical and epidemiological research, where simplicity and actionability often outweigh granular precision. Diverse applications from heart rate thresholds in pulmonary disease \citep{Wells_2000} to medication adherence in atrial fibrillation patients \citep{safari_helisaz_2024, salmasi_safari_2024} require statistically sound approaches to identify optimal cut-points for discretization. While thresholds such as `heart rate $> 100 bpm$' or `medication adherence $> 80\%$' are entrenched in clinical practice, they are often derived from expert consensus or arbitrary quantiles, risking suboptimal power or missed biological signals. Modern high-dimensional datasets demand data-driven cut-point detection methods that balance interpretability with predictive accuracy.

Current approaches to cut-point detection suffer from several key limitations. Methods based on multiple testing (e.g., \citealt{Bland_Altman_1995, Lausen_Schumacher_1992, Rota_2015}) are computationally inefficient, often restricting analysis to a small set of candidate thresholds and potentially missing optimal cut-points. Other approaches focus solely on either single-predictor settings (e.g., \citealt{OBrien_2004}), allowing single cut-point per feature (e.g., \citealt{Icuma_2018}), or detecting cut-points for a single feature at a time \citep{Leblanc_1993, Motzer_1999, Chang_2019}, rendering them inadequate for high-dimensional data where joint selection and thresholding of multiple variables is essential.

To our knowledge, the only existing scalable statistical approach for data-driven cut-point detection for survival analysis is the binacox method by \citet{bussy_2022}. This method addresses cut-point identification in high-dimensional Cox models, particularly relevant for medical and genetic studies where multiple cut-points per feature are often needed (e.g., \citealt{Cheang_2009}). binacox employs one-hot encoding combined with the binarsity (total-variation [TV]) penalty along with a linear constraint \citep{alaya_2019} for cut-points detection of continuous features.

Despite its strengths, the binacox method has some limitations that warrant consideration. A primary concern is its reliance on TV regularization forces piecewise‑constant solutions that assume abrupt risk changes, making it ill-suited for detecting gradual biological effects. Furthermore, the method's stability is contingent on having sufficient sample sizes within each created bin, making it sensitive to unevenly distributed covariates and prone to instability in sparsely sampled regions. Finally, the binacox model's focus on local effect changes causes it to neglect the overall trend of a covariate's influence, particularly near its boundaries. This can result in the identification of suboptimal cut-points that do not accurately represent the covariate's broader relationship with the outcome.

In this work, we introduce biniLasso, a new approach to cut-point detection that addresses key limitations of existing methods like binacox. biniLasso employs cumulative binarization, a paradigm shift that more accurately captures the underlying effects of continuous covariates. This framework not only improves cut-point detection but also naturally aligns with clinical interpretation, where risk thresholds (e.g., `above/below a critical value') are more meaningful than discrete categories. Additionally, we propose miniLasso by integrating the recently developed uniLasso method, which enforces sparsity while preserving the sign of univariate model coefficients and their magnitude \citep{unilasso_2025}. This dual innovation, cumulative binarization plus uniLasso, ensures both interpretability (via threshold-aligned effects) and statistical efficiency (via sparse, univariate-consistent estimates). We validate the effectiveness of biniLasso and miniLasso through an extensive simulation study as well as illustrating their practical utility by applying them to three high-dimensional cancer genomics datasets.

The remainder of this paper is structured as follows: Section~2 details our proposed biniLasso and miniLasso approaches.
Sections~3 and 4 report the results of a comprehensive simulation study and a case study, respectively. Finally, Section~5 presents the discussion and conclusions.

\section{Method}
\label{s:method}

We present the biniLasso approach, including its binarization modification step, estimation procedure, and implementation details. We adopt the usual notation in the framework of survival analysis \citep{Andersen_2012} to describe the variables and models
Specifically, let \( (X_i, Z_i, \Delta_i) \in \prod_{j=1}^{p}[a_j, b_j]^p \times \mathbb{R}^+ \times \{0, 1\} \), for \( i = 1, \dots, n \) be the observed triple of the $i^{th}$ observation in the independent and identically distributed sample of size $n$, where $X_i$ is the vector of covariates, $Z_i$ is the time to event (possibly right-censored), $\Delta_i$ is the censoring indicator (1 when $Z_i$ is fully observed and 0 when is right-censored), and the boundary values \( a_j \) and \( b_j \) of the $j^{th}$ predictor may extend to \( -\infty \) and \( \infty \), respectively. If population-level minimum and maximum values of the predictors were available, one could normalize the predictors accordingly, simplifying the notation by setting \( a_j \equiv 0 \) and \( b_j \equiv 1 \).

The Cox proportional hazards model \citep{cox_1972} is used to describe the relationship between the hazard function and predictor variables, modelled as $\lambda(t | X_i) = \lambda_0(t) \exp\left( f(X_i) \right)$, where $\lambda_0(t)$ is the baseline hazard function, and $f(\cdot)$ , as a linear function with respect to the regression coefficients, quantifies the relationship between the covariates $X_i$ and the outcome hazard. The primary objective is to estimate the function $f(\cdot)$.

\subsection{Model construction}

We begin by transforming the covariates into a binarized matrix $X^B$, where each continuous variable is encoded into a set of binary dummy variables. This encoding expands the original design matrix of $X$ with $p$ columns into $p+d$ columns, possibly $d >> p$, where the $j^{th}$ continuous feature is replaced by $d_j +1 \ge 2$ binary columns $X_{.,j,1}^B, \dots, X_{.,j,d_j+1}^B$, and $d = \sum_{j=1}^p d_j$.

The intervals $I_{j,1}, \dots, I_{j, d_j+1}$ are subsets of the range of the $j^{th}$ continuous covariate such that their union equals to the range of the covariate and for each observation $i=1, \dots, n$ and for each feature $j$, the binarized covariate $X_{i,j,l}^B$ is defined as
$X_{i,j,l}^B=\left\{\begin{array}{lll}
                1 & \textnormal{if} & X_{i,j} \in I_{j,l} \\
                0 & \textnormal{} & \textnormal{otherwise}
            \end{array}\right.$.
Depending on the purpose of categorization, the intervals $I_{j,l}$'s can be constructed differently. For instance, in binacox, the intervals $I_{j,l}$'s were a partition of the range of the $j^{th}$ continuous covariate. Therefore, the resulting binarized matrix $X^B$ was a sparse matrix of one-hot encoded versions of the original continuous covariates. We will propose a different method of constructing the intervals $I_{j,l}$'s in the next section.
The Cox linear predictor function of $f_{\beta}(\cdot)$ can then be represented in terms of such binarized covariates as follows:
\begin{equation} \label{eq:fBeta}
\begin{split}
f_{\beta}(X_i) &= \boldsymbol{\beta}^T \boldsymbol{X}_i^B \\
&= \sum_{j=1}^{p} f_{\boldsymbol{\beta}_{j,.}}(X_{i,j}) \\
&= \sum_{j=1}^p \sum_{l=1}^{d_j+1} \beta_{j,l} \boldsymbol{1}(X_{i,j} \in I_{j,l})
\end{split}
\end{equation}
where the vector of coefficients $\boldsymbol{\beta}$ is given by $\boldsymbol{\beta} = (\boldsymbol{\beta}_1^T, \dots, \boldsymbol{\beta}_p^T) ^ T = (\beta_{1,1}, \dots, \beta_{1,d_1+1}, \dots, \beta_{p,1}, \dots, \beta_{p, d_p+1}) ^ T$
Then, the scaled negative log-partial likelihood function of the Cox model is:
\begin{equation} \label{NLL}
\l_n(f_{\beta}) = -\frac{1}{n} \sum_{i=1}^n \Delta_i \left\{ f_{\beta}(X_i) - \log \sum_{i^`: Z_{i^`} \ge Z_i} \exp\left(f_{\beta}(X_{i^`})\right) \right\}
\end{equation}

\subsection{Cumulative binarization}
Our method employs a cumulative binarization technique to represent continuous covariates. Unlike standard one-hot encoding, which creates a separate indicator for each discrete bin, our approach generates a series of dummy variables for the splitting points (cut-points) themselves. For each splitting point, the corresponding dummy variable indicates whether an observation's value is greater than that point. This creates a nested structure, where the intervals for larger splitting points are contained within those for smaller ones, effectively capturing a cumulative effect across the covariate's range.

Let \( X^{CB} \) represent the binarized matrix with \( p + d \) columns, where continuous features are cumulatively multi-hot encoded. Unlike binacox, we relax the assumption of known population-level minimum and maximum values for the predictors \( X \) to rescale them to $[0, 1]$. Continuous predictors can still be standardized (or normalized) based on the sample data before fitting the model, as is common when penalty terms involve the absolute values of predictor coefficients. Importantly, these preprocessing steps do not require access to population-level data.

For the \( j^{th} \) feature with \( d_j + 1 \) cumulative binarized columns, we define strictly increasing endpoints \( \mu_{j,l}\), where \(l = 0, \dots, d_j \) (potential cut-points). These endpoints create nested, decreasing intervals \( I^c_{j,l} = (\mu_{j,l}, b_j] \) for \( l = 0, \dots, d_j \), with \( I^c_{j,0} = (a_j, b_j] \). For the $i^{th}$ observation, the $j^{th}$ predictor, and the $l^{th}$ interval, the cumulative binarized variable \( X_{i,j,l}^{CB} \) is then defined as
$X_{i,j,l}^{CB} = \left\{\begin{array}{lll}
                1 & \textnormal{if} & X_{i,j} \in I^c_{j,l}, \\
                0 & \textnormal{} & \textnormal{otherwise}
            \end{array}\right.$.
Additionally, let $\boldsymbol{X}_{j,l}^{CB}$ be the vector of the $l^{th}$ cumulative binarized column corresponds to the $j^{th}$ feature.

The rationale behind this cumulative binarization is to facilitate the interpretation of the $j^{th}$ continuous covariate at each cut-point by comparing ``lower versus all higher values": values in $(a_j, \mu_{j,l}]$ (``lower") vs values in the complement of that interval (``all higher"). 
This enables the direct estimation of the effect size for such low/high comparisons. By simultaneously including multiple cumulative binarized covariates, we can comprehensively perform comparisons across different intervals of the continuous covariate, providing a richer understanding of the variable’s relationship to the outcome. While these cumulative binarized features are (positively) correlated, unlike standard one-hot encoding, the resulting design matrix here is not full-rank.

The cumulative binarization framework is particularly well-suited for detecting risk thresholds, identifying cut-points such that values in the interval up to the threshold exhibit a risk profile similar to the variable’s minimum (or maximum) value. To anchor this interpretation, boundary indicators are included as unpenalized terms, allowing regularization to isolate distinct risk intervals within the interior of the covariate range. For a full technical discussion of this application, including model specification and motivation, see Supplementary Materials (SM) Section~1.

\subsection{Estimation procedure}

For cut-points detection of continuous covariates, regularization is a natural solution as we are aiming to select optimal cut-points from a set of candidate cut-points for each covariate. A constrained total variance penalty term (a group Fused Lasso like penalty) was employed in binacox.
We propose a different regularization problem based on cumulative binarization. Specifically, for each binarized feature $\boldsymbol{X}_{j,l}^{CB}$, there corresponds a parameter $\beta^*_{j,l}$. The vector associated with the binarization of the $j^{th}$ feature is denoted by $\boldsymbol{\beta}^*_j = (\beta^*_{j,1}, \dots, \beta^*_{j,d_j})^T$. Each parameter $\beta^*_{j,l}$ is linked to a corresponding cut-point $\mu_{j,l}$, thus the parameter vector $\boldsymbol{\beta}^*_j$ corresponds to the cut-point vector $\boldsymbol{\mu}_j = (\mu_{j,1}, \dots, \mu_{j,d_j} )^T$. Using this parameterization, a candidate function for the estimation of $f$, denoted as $f_{\boldsymbol{\beta}^*}(\boldsymbol{X}_i)$, can be expressed similarly as in \ref{eq:fBeta}, in which, the full parameter vectors of size $p+d$ and $d$, respectively, are obtained by concatenating the vectors $\boldsymbol{\beta}^*_j$ and $\boldsymbol{\mu}_j$, similar to the formulation previously used.

To estimate the parameter $\beta^*$, we apply a weighted Lasso penalized Cox partial likelihood approach. The optimization problem is defined as:
\begin{equation} \label{opt_bini}
\hat{\boldsymbol{\beta}}^* = argmin_{\boldsymbol{\beta}^*} \left\{ \l_n(f_{\boldsymbol{\beta}^*}) + \sum_{j=1}^p \left(\sum_{l=1}^{d_j+1} w^*_{j,l} |\beta^*_{j,l}| \right) \right\}
\end{equation}
where
\begin{equation*}
\begin{split}
\l_n(f_{{\beta}^*}) = -\frac{1}{n} \sum_{i=1}^n \Delta_i &\left\{ f_{\beta^*}(X_i) \right.\\
&- \left.\log \sum_{i^`: Z_{i^`} \ge Z_i} \exp(f_{\beta^*}(X_{i^`})) \right\}
\end{split}
\end{equation*}
While assigning different weights \( w^*_{j,l} \) to each parameter \( \beta^*_{j,l} \) as additional tuning constants enhances model flexibility, it comes at the cost of increased computational complexity during model fitting. As an alternative, these weights can be specified using strategies similar to those employed in group and adaptive Lasso techniques (e.g., cross validation).

\subsection{miniLasso}

While the Lasso has been widely adopted in many applications, it suffers from well-known limitations, particularly its sensitivity to correlated predictors. This issue is especially relevant in our context, as cumulative binarization inherently creates correlated features. To address this challenge while maintaining both sparsity and interpretability, we incorporate uniLasso \citep{unilasso_2025}, a novel two-stage regularized regression approach. The uniLasso procedure consists of three key steps: 1. First, fit univariate Cox models for each individual indicator variable of each binarized feature, generating linear predictor functions $\hat{\eta}_{j,l}(x_{i,j,l})$ for $j=1,\dots,p$ and $l=1,\dots,d_j$, 2. Compute leave-one-out (LOO) predictions $\hat{\eta}_{j,l}^{-i}$ for all $n$ observations and each $(j,l)$ pair, and 3. Fit a non-negative Lasso Cox model using these LOO predictions as features.
This two-stage approach provides important theoretical and practical advantages. First, the univariate Cox models (Step 1) and non-negative Lasso Cox model (Step 3) together ensure sign consistency, i.e., the estimated coefficients in the multivariable model preserve the direction of effects from their univariate counterparts. Second, by design, this approach maintains comparable coefficient magnitudes between univariate and multivariable models without requiring feature standardization in Step 3. Finally, using leave-one-out predictions (Step 2) enhances the robustness and predictive performance of the final multivariate model by reducing overfitting \citep{unilasso_2025}.
This leads to the modified optimization problem $\hat{\boldsymbol{\theta}} = argmin_{\boldsymbol{\theta}} \left\{ \l_n(f_{\boldsymbol{\theta}}) + \sum_{j=1}^p \left(\sum_{l=1}^{d_j+1} w^{**}_{j,l} |\theta_{j,l}| \right) \right\}$, subject to $\theta_{j,l} \ge 0$ for all $j$ and $l$, where $f_{\boldsymbol{\theta}}(\boldsymbol{X}_i) = \sum_{j=1}^p \sum_{l=1}^{d_j} \theta_{j,l} \hat{\eta}_{j,l}^{-i}(x_{i,j,l})$, and $w^{**}_{j,l}$ are adaptive weights. The additional sparsity from this approach motivates our designation of this method as miniLasso.

\subsection{Theoretical properties of the cumulative basis}

The adoption of cumulative binarization in biniLasso fundamentally alters the estimation problem compared to binacox. This transition is not a mere data-preprocessing modification, but a formal bijective linear transformation (change of basis) that maps the complex, non-smooth TV penalized space of binacox onto a strictly separable, unconstrained $\ell_1$ geometry. This basis transformation diagonalizes the penalty operator, eliminating the need for the coupled TV penalty. Consequently, the optimization problem simplifies to a standard, unconstrained $\ell_1$-penalized Cox model that directly inherits the well-established non-asymptotic oracle properties and variable-selection consistency of the classical Lasso (e.g., \citet{Tibshirani_1997, Huang_2013}). Crucially, by isolating threshold effects into decoupled, univariate parameters, this cumulative framework serves as the necessary mathematical foundation for enforcing the sign-preserving univariate consistency of our sparse miniLasso variant. An analytical exposition of these mathematical and algorithmic shifts is provided in Section 2 of the SM.

Beyond these structural and computational advantages, the cumulative transformation grounds the high-dimensional cut-point detection problem in well-established theoretical domains. While cumulative indicators are inherently (possibly highly) correlated, a condition that traditionally complicates exact variable selection consistency in generic Lasso applications, this specific highly-ordered, lower-triangular design matrix has been rigorously validated in the multiple change-point estimation literature. Research demonstrates that mapping a one-dimensional TV penalty onto a cumulative step-basis satisfies a specialized Restricted Eigenvalue condition \citep{Rinaldo_2009, Harchaoui_2010}. Consequently, the standard $\ell_1$-penalized Cox formulation in biniLasso guarantees robust non-asymptotic estimation rates and consistent threshold recovery despite the inherent correlation in the transformed predictors. A detailed discussion of these oracle properties under our specific correlated design is detailed in Section 2 of the SM.

\subsection{Limit number of cut-points} \label{fixedcuts}

For enhanced interpretability, it is often desirable to categorize continuous covariates using a limited number of cut-points. One approach to constrain the number of cut-points for each predictor is to modify the penalty term in \eqref{opt_bini} by assigning predictor-specific weights, $w^*_{j}$, $\sum_{j=1}^p w^*_{j} \left(\sum_{l=1}^{d_j+1} |\beta^*_{j,l}| \right)$,
Here, $w^*_{j}$ is a weight calibrated to yield a specific number of non-zero coefficients, $\beta^*_{j,l}$, for the $j^{th}$ predictor. While this method can control regularization in low dimension, its practical application is problematic, especially in high-dimensional settings. Iteratively adjusting these weights to force a pre-specified number of non-zero coefficients is computationally expensive, may not converge, and can be unstable. This difficulty arises because the task is fundamentally a combinatorial (non-convex) subset selection problem, whereas the Lasso framework is designed for convex optimization governed by a single global tuning parameter, $\lambda$.

We propose a computationally efficient, two-step procedure to select at most $m$ cut-points for each predictor. First, for each continuous predictor, we fit a separate Lasso-penalized Cox model (or a GLM) to its binarized features over a fine grid of $\lambda$ values. This step is performed independently for each predictor. By examining the resulting coefficient path, we identify and retain the top $m$ most influential cut-points (i.e., the binarized features that enter the model earliest or have the largest coefficients). This initial screening effectively ranks and filters the potential cut-points for each predictor based on its individual predictive power. Next, we construct a single, final model. The feature set for this model comprises only the top $m$ binarized features for each predictor, as selected in Step 1. We then fit a standard Lasso Cox model to this combined set of features. The final selection of cut-points is determined by the non-zero coefficients at the optimal $\lambda$ value, chosen via CV.

This procedure ensures that no more than $m$ cut-points are selected for any given predictor. If the goal is to select exactly $m$ cut-points for any predictor that remains in the model, the second step can be modified. Instead of a standard Lasso Cox model, a group-Lasso Cox model is fitted. The $m$ selected binarized features for each original predictor are defined as a group. This forces the model to either include all $m$ features for a predictor or exclude the predictor entirely, thus ensuring that any selected predictor has precisely $m$ cut-points.

The two-step procedure is designed to handle multiple competing predictors. However, for simpler cases involving the categorization of only one or two continuous variables, a direct, one-step approach is sufficient. This simplified method involves fitting a single, comprehensive Lasso Cox model (or more broadly a GLM) with the binarized features of all relevant predictors included at once. The top $m$ cut-points for each predictor are then chosen directly from the coefficient path of this global model.

\section{Simulation study}
\label{s:simulation}

\subsection{Simulation designs, benchmarks, and evaluation metrics}
\begin{figure*}[hbt!]
\centering
        \includegraphics[scale=0.3]{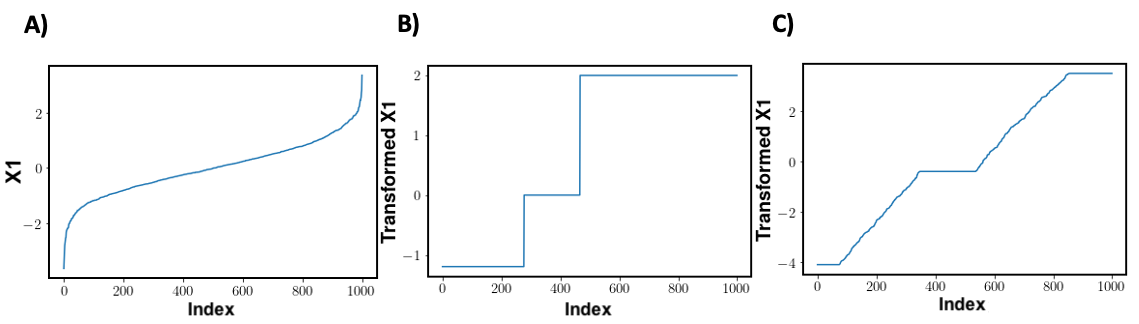}
    \caption{(A) Original continuous predictor $X_1$ used as input. (B) True threshold relationship in Scenarios 1, 2, and 4, with two true cut‑points. (C) Gradual ``cut‑region'' relationship in Scenario 3.}
    \label{fig:allScens}
\end{figure*}

We conducted a comprehensive simulation study to evaluate the performance of biniLasso and miniLasso, against the benchmark method binacox, the only existing regularization-based approach suitable for high-dimensional cut-point detection. We simulated survival data under Cox proportional hazards models using two underlying relationships: a step-function with sharp cut-points (Figure~\ref{fig:allScens}B) and a gradual ``cut-region'' where hazard changes linearly within specified intervals (Figure~\ref{fig:allScens}C), while only the predictors original form (Figure 1A) were available to each method.

Our evaluation spanned four distinct scenarios. Scenarios 1 and 2 served as benchmarks, replicating earlier designs for direct comparison: the first uses two predictors with true cut‑points, while the second extends to high dimensions (2-100 predictors) with $20\%$ sparsity. Scenario 3 introduces a realistic challenge by replacing sharp thresholds with smooth cut‑regions. Scenario 4 assesses performance under an interpretability constraint, forcing exactly two cut‑points per predictor to reflect clinical preference for simple categorization.

For the main simulation study, continuous covariates were partitioned using $d_j = 50$ candidate bins, with candidate cut-point placements determined by the empirical quantiles of each respective covariate. We assessed each method using multiple criteria: computation time, average number of estimated cut-points per predictor (structural stability), overall model performance (predictive stability) via Akaike’s Information Criterion (AIC) and the Integrated Brier Score (IBS), and where true cut-points existed, detection accuracy measured by the enhanced Hausdorff set-distancee between estimated and true cut-points (threshold stability). All main simulations were conducted over five sample sizes (300, 500, 1000, 2000, and 4000) and repeated 5000 times for stability. Detailed data-generating mechanisms and metric definitions are provided in SM Sections 4.1 and 4.2

To evaluate the robustness of our models to number of cut-points candidates and their placement choices, we incorporate targeted sensitivity analyses. First, we assess the impact of grid resolution by varying the number of bins ($d_j$) from 10 to 70. This tests the expected trade-off between the theoretical precision of cut-point detection (which favors a denser grid) and statistical stability (which favors fewer bins to maintain adequate sample sizes per interval). Second, we compare the default quantile-based partitioning against equal-interval (fixed) binning to determine the sensitivity of threshold recovery to the placement strategy. These analyses were evaluated at a sample size of $n=4000$ and were restricted to simulation scenarios 1 (true cut-points existed) and 3 (true continuous underlying relationship).

All statistical analyses and visualizations were performed using the R programming language (R Core Team, 2023). To facilitate the methods proposed in this paper, we developed the biniLasso R package available on GitHub at \url{https://github.com/ab-sa/biniLasso}, as well as the source code to reproduce all simulation studies and the case study analysis is publicly available on GitHub at \url{https://github.com/ab-sa/biniLasso-paper}. More details are provided in SM section 3.

\subsection{Simulation results}
\begin{figure*}[!htbp]
\centering
        \includegraphics[scale=0.4]{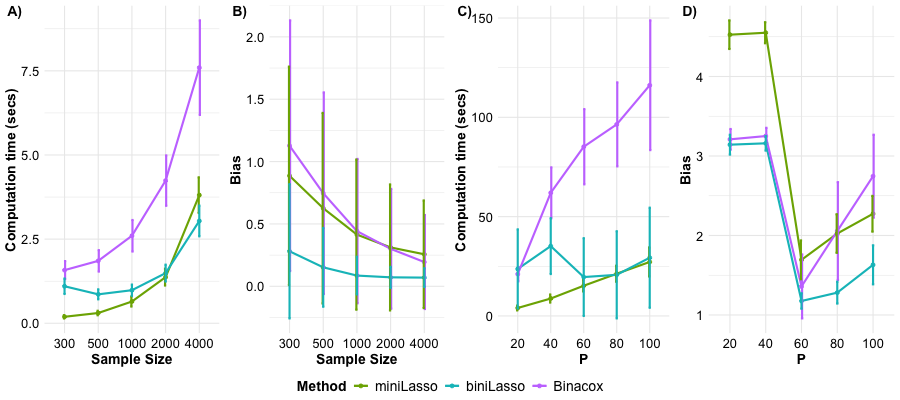}
    \caption{Results for benchmark Scenarios 1 and 2. For Scenario 1: average computation time (A) and bias in estimated cut‑points (B) across $n$'s. For Scenario 2: average computation time (C) and bias (D) across $P$'s. Results are shown for biniLasso (blue), miniLasso (green), and binacox (purple). Vertical bars represent ± 1 SD over 5000 simulations.}
    \label{fig:sc1_t_perf}
\end{figure*}

Figure~\ref{fig:sc1_t_perf} compares the computational efficiency and the estimated cut-points bias of biniLasso (blue), miniLasso (green), and binacox (purple) across varying sample sizes under Scenario 1 (panels A and B) and varying number of predictors under Scenario 2 (panels C and D). Both biniLasso variants demonstrated substantially faster computation times than binacox, with speed improvements ranging from 2-fold (larger $n$'s and smaller $P$'s) to 8-fold (smaller $n$'s or larger $P$'s). Though biniLasso appeared smaller average bias compare to both miniLasso and binacox under Scenario 1, but the differences were not significant. Under Scenario 2, however, biniLasso had smaller bias for most values of $P$. No signifiant differences found in the AIC, IBS, and number of estimated cut-points of the different methods under Scenario 1 (Figure~S.1 in SM). Under Scenario 2, biniLasso has a lower AIC average for larger $P$'s but no significant difference in IBS across different methods (Figure~S.2 in SM).

\begin{figure*}[!htbp]
\centering
        \includegraphics[scale=0.38]{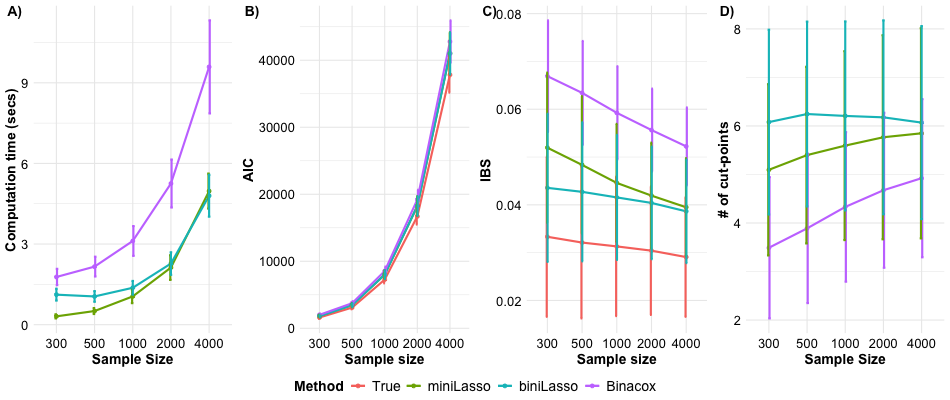}
    \caption{Results for Scenario 3 (no true cut-points). Average computing time (A), AIC (B), IBS (C), and number of estimated cut-points for $X_1$ (D) for biniLasso (blue), miniLasso (green), binacox (purple), and the true continuous model (red). Results are shown across $n$'s and vertical bars represent ± 1 SD over 5000 simulations.}
    \label{fig:Sc3_all}
\end{figure*}

Figure~\ref{fig:Sc3_all} compares computational efficiency (A) and model fit (B–D) under Scenario 3, where no true cut-points exist. Both biniLasso variants maintained a substantial speed advantage over binacox (A). In terms of model quality, all discretization methods produced AIC values comparable to the true continuous model (B), indicating they successfully captured the underlying relationships despite the absence of thresholds. biniLasso also achieved prediction accuracy (IBS) on par with the true model, while binacox showed significantly higher IBS across all sample sizes (C). Model complexity, measured by the average number of estimated cut-points, was similar across methods (D).

In Scenario 4 (two continuous predictors), we used the one‑step approach (Section \ref{fixedcuts}\textcolor{red}{2.6}) to select exactly two cut‑points per predictor for biniLasso and miniLasso. Both variants again computed faster than binacox. By design, all methods produced equally complex models; however, binacox’s single tuning parameter could not guarantee exactly two cut‑points. Therefore, comparisons were restricted to the approximately $10\%$ of runs where binacox fortuitously returned exactly two cut‑points (Table S.1). In contrast, biniLasso’s feature‑specific tuning allowed precise control. As expected, the true continuous model outperformed all categorization methods on both AIC and IBS, especially at larger sample sizes. Among the categorized models, all had similar average AIC, but biniLasso achieved the lowest IBS (closest to the true model) while miniLasso and binacox performed similarly. These results are detailed in Figure~S.3.

\subsection{Simulation sensitivity analysis}

The sensitivity analyses across both Scenarios 1 and 3 settings demonstrated that quantile-based partitioning looked more robust than fixed-interval binning, particularly at lower grid resolutions (Figures S.4 and S.5, respectively). While fixed-interval binning yielded higher discrepancy and more severe underestimation with fewer than 30 bins, both placement strategies seemed to converge in performance at higher resolutions, with predictive accuracy (AIC, IBS) and discrepancy showing diminishing returns after 30 to 50 candidate bins. When true discrete cut-points exist, miniLasso seemed to maintaining stable cut-point recovery even as the grid becomes denser, whereas the biniLasso tended to slightly overestimate. Conversely, when approximating continuous relationships, both algorithms seemed to estimate more cut-points as resolution increases to capture the underlying gradient. Across all settings, computation time scaled linearly. See Section 5.6 of SM.

\section{Case study} \label{s:application}
\subsection{Datasets}
To evaluate the practical performance of our method, we applied biniLasso, miniLasso, and binacox to three The Cancer Genome Atlas (TCGA) cancer datasets: breast invasive carcinoma (BRCA, $n=1,231$), glioblastoma multiforme (GBM. $n=391$), and kidney renal clear cell carcinoma (KIRC, $n=614$), each with gene expression (FPKM values for 60,660 genes) and survival outcomes.

Due to high dimensionality, we first performed gene screening by fitting univariate Cox models and ranking genes by AIC and IBS. The top 50 genes from each metric were retained (up to 100 unique genes per dataset). This continuous‑Cox screening step was computationally efficient and effectively enriched for genes with detectable survival associations, including those with threshold effects; sensitivity analysis confirmed strong concordance with cut‑point‑based log‑rank tests (Figure~S.6 in SM). Selected genes were then standardized before applying the cut‑point detection methods. Dataset summaries after screening are provided in Table S.1.

\subsection{Estimated cut-points and model performance}
We applied biniLasso, miniLasso, and binacox to each dataset after the screening step. In line with the simulation study, we set 50 candidate bins (at quantiles) for all predictors (i.e., $d_j=50$ for all $j$). For biniLasso and miniLasso, we additionally employed the two-step approach described in Section~2.6, restricting each gene to at most two cut‑points. Table~S.4 reports the estimated cut‑points for the KIRC dataset. biniLasso, miniLasso, and binacox identified a single cut‑point for 20, 15, and 21 genes, two cut‑points for 5, 2, and 1 genes, and three cut‑points for 2, 1, and 0 genes, respectively. In the BRCA dataset (Table~S.5), cut‑points were detected for 36, 30, and 29 genes by biniLasso, miniLasso, and binacox; in the GBM dataset (Table~S.6), the corresponding numbers were 43, 34, and 29 genes. The cut‑point sets estimated by biniLasso and miniLasso largely overlapped, though miniLasso produced somewhat sparser selections. Greater discrepancies were observed between binacox and both variants of biniLasso.

To assess the impact of cut-point detection on model performance, we fitted Cox models using categorized predictors based on each method's detected cut-points (retaining genes with at least one cut-point). As a continuous benchmark, we also fitted a Cox-based Generalized Additive Model (CGAM) with smoothing splines for all selected genes. Performance was evaluated via 10-fold cross-validation (CV), with results summarized in Table~\ref{tab:metrics}. We report the average relative AIC, IBS, and Concordance Index (C-index), along with their SDs, compared to the CGAM benchmark.

Contrary to expectations, biniLasso and miniLasso occasionally matched or slightly exceeded the performance of the flexible CGAM benchmark. Among the three binarization methods, biniLasso consistently outperformed binacox. biniLasso also generally (but not significantly) outperformed miniLasso, with only two minor exceptions. Notably, miniLasso achieved this comparable performance while detecting fewer cut-points than biniLasso. binacox identified the fewest cut-points overall but exhibited the highest variation in cut-point estimates across CV folds.
\begin{table*}[h!]
   \caption{Cross-validated performance metrics relative to CGAM for Cox models using categorized predictors derived from biniLasso, miniLasso, and binacox cut-points for BRCA, GBM, and KIRC datasets. Reported values are mean (SD) of relative AIC, IBS, and C-index across 10 folds.}
   \label{tab:metrics}
   \small
   \centering
   \begin{threeparttable}
   \begin{tabular}{llccc}
   \textbf{Dataset} & \textbf{Metric} & \textbf{biniLasso} & \textbf{miniLasso} & \textbf{binacox} \\ 
   \hline
   BRCA & Relative AIC\tnote{1} (SD) & \textbf{0.952 (0.010)} & 0.967 (0.007) & 1.00 (0.019) \\
   BRCA & Relative IBS\tnote{2} (SD) & \textbf{1.130 (0.647)} & 1.140 (0.379) & 1.273 (0.425) \\
   BRCA & Relative C-index\tnote{3} (SD) & 0.972 (0.129) & \textbf{0.987 (0.142)} & 0.875 (0.154) \\
   BRCA & No. of cut-points (SD) & 39.8 (8.664) & 33.0 (4.497) & 16.1 (10.898) \\
   \hline
   GBM & Relative AIC (SD) & \textbf{0.955 (0.013)} & 0.971 (0.007) & 1.010 (0.017) \\
   GBM & Relative IBS (SD) & \textbf{0.920 (0.168)} & 0.946 (0.189) & 1.017 (0.175) \\
   GBM & Relative C-index (SD) & \textbf{1.068 (0.089)} & 1.054 (0.090) & 0.948 (0.088) \\
   GBM & No. of cut-points (SD) & 51.7 (11.605) & 44.4 (3.658) & 16.2 (9.739) \\
   \hline
   KIRC & Relative AIC (SD) & 0.964 (0.008) & \textbf{0.961 (0.007)} & 0.982 (0.006) \\   
   KIRC & Relative IBS (SD) & \textbf{0.963 (0.161)} & 0.980 (0.177) & 1.007 (0.172) \\
   KIRC & Relative C-index (SD) & \textbf{1.035 (0.079)} & 1.030 (0.083) & 0.998 (0.093) \\
   KIRC & No. of cut-points (SD) & 36.6 (4.526) & 27.8 (4.662) & 33.2 (10.942) \\
   \hline \\
   \end{tabular}
   \begin{tablenotes}
    \item[1] Akaike’s Information Criterion
    \item[2] Integrated Brier Score
    \item[3] Concordance Index
  \end{tablenotes}
\end{threeparttable}
\end{table*}

\begin{figure*}
\centering
        \includegraphics[scale=0.45]{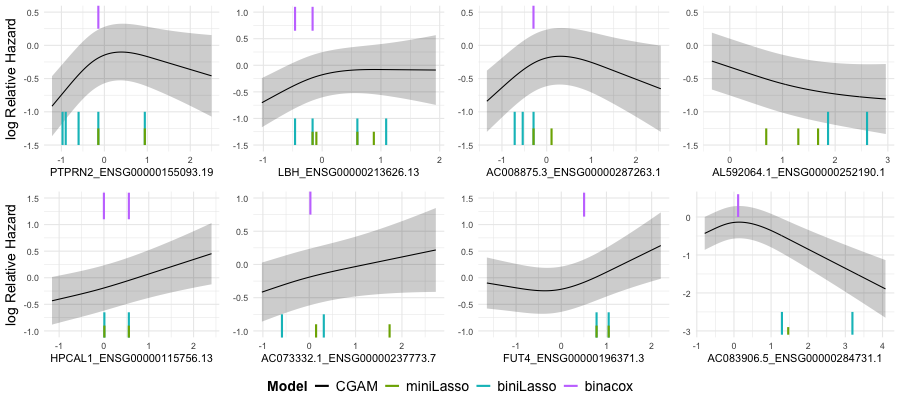}
    \caption{The detected cut-points for the selected 8 genes with the highest number of cut-points across all three methods versus log relative hazard from a CGAM smooth fit in GBM data.}
    \label{fig:app_gbm}
\end{figure*}

Figure~\ref{fig:app_gbm} illustrates the location and nature of the detected cut-points for the selected 8 genes with the highest number of cut-points across all three methods in GBM data. For each, we plotted the estimated log relative hazard (black curve) from a CGAM smooth fit and overlaid the cut-points detected by each method (biniLasso in blue, miniLasso in green, and binacox in purple). A clear methodological pattern emerged: cut-points identified by binacox tended to align with inflection points where the hazard trend changes direction. In contrast, biniLasso and miniLasso tended to place cut-points at regions where the slope of the hazard function is steepest - locations where predictions change most rapidly. Additionally, when the hazard function changes gradually, binacox failed to detect any cut-points (e.g., gene AL592064.1). Figure~S.7 provides similar plots for the BRCA and KIRC data.

We fitted Cox proportional-hazards models using cut-points selected by the biniLasso and miniLasso methods. Both were constrained to a maximum of two cut-points per gene, identified using the two-step approach from Section~2.6. As summarized in (Table S.3 in SM), the predictive performance of the two methods was nearly identical. However, miniLasso consistently produced more parsimonious models, selecting fewer cut-points both within individual genes and across the entire dataset.

\section{Discussion}
\label{s:discussion}

In this work, we introduce biniLasso and its sparse variant miniLasso for efficient, multi-cut-point detection in high-dimensional Cox models. Simulations show that both methods outperform binacox computationally (2–8× faster) while matching or improving accuracy in cut-point recovery and model fit. Standard biniLasso detects multiple cut-points per feature for complex relationships; miniLasso enforces sparser, more interpretable models. When the number of cut-points is constrained - a common practical scenario - predictive performance between the two methods becomes nearly identical, making miniLasso a parsimonious and performance-preserving choice. These advantages are validated in applications to three high-dimensional genomic cancer datasets, demonstrating practical utility in biomedical research.

The cumulative binarization approach used in biniLasso addresses key limitations of conventional binarization methods, which are often sensitive to arbitrary interval boundaries and can produce small, underpowered bins. By using nested intervals, biniLasso enables more flexible and stable cut-point estimation, mitigating residual confounding and loss of precision, especially near the extremes of the data range. The choice between standard biniLasso and its sparse variant miniLasso depends on the analytical goal: biniLasso is ideal for exploratory discovery where multiple cut-points capture complex relationships, while miniLasso provides a parsimonious, interpretable model well-suited for developing simplified clinical risk categories. Importantly, both variants are readily extensible to other GLM families (binary, count, continuous outcomes) via the glmnet framework, broadening their applicability beyond survival analysis.

The methodological foundation of biniLasso rests on pairing cumulative binarization with an ordinary $\ell_1$ penalty. Unlike one-hot frameworks that necessitate computationally heavy, constrained Fused Lasso optimization, this cumulative basis maps the problem into a strictly separable $\ell_1$ geometry. This transformation diagonalizes the penalty, unlocking highly scalable coordinate descent and inheriting robust change-point theoretical guarantees despite structured correlation. While alternative penalties are often applied to correlated data, they undermine this foundation: the ``grouping effect'' of Elastic Net erroneously smears sharp clinical thresholds across adjacent bins, and non-convex penalties (SCAD, MCP) sever the TV isomorphism. Thus, the $\ell_1$ cumulative framework uniquely delivers the exact sparsity, convexity, and theoretical rigor required for distinct cut-point detection and our sign-constrained miniLasso architecture.

Our study has some limitations. The performance of both biniLasso variants depends on the pre-specified set of cut-point candidates, which may not always reflect the true underlying structure of the data. Additionally, the method assumes that cut-point-based modeling adequately captures predictor-outcome relationships, which may not hold for all data types. Future work could explore adaptive strategies for selecting cut-points and extend biniLasso to accommodate more complex data structures including at the presence of interactions. A particularly promising direction would be developing a method to detect cut-points directly from continuous covariates without relying on pre-specified candidates. Such an approach could improve both efficiency and performance by objectively deriving data-driven cut-points, eliminating the subjectivity inherent in manual candidate selection.

How can this approach be utilized in real-world applications? We strongly advocate for first identifying the optimal relationship between predictors and the risk of outcomes, irrespective of its immediate clinical interpretability. This can be achieved using CGAMs, which leverages smoothing spline-based machine learning algorithms \citep{Hastie_Tibshirani_1990, Bender_etal_2018}. In this approach, continuous predictors are included in the Cox model, and smoothing splines are applied to flexibly estimate potentially non-linear, data-driven relationships between predictors and outcomes. Then, for generating clinically actionable insights, one can categorize continuous predictors based on identified cut-points by biniLasso (or miniLasso) and then fitting a standard Cox proportional hazards model using the categorized predictors instead. This approach balances interpretability and performance, making it more suitable for practical applications. To evaluate the trade-off between interpretability and predictive accuracy, model performance metrics such as AIC or IBS can be used to compare the simplified Cox model with the more flexible CGAM. Notably, if the true biological mechanism exhibits threshold or switch-like effects rather than smooth continuous gradients, the categorized model may actually outperform the CGAM by avoiding the over-parameterization inherent to flexible splines. Additionally, bootstrap resampling or stability selection can be used to quantify estimated cut-points uncertainty.

In conclusion, biniLasso and miniLasso represents a significant advancement in the analysis of high-dimensional survival data, offering a computationally efficient and interpretable approach for identifying multiple cut-points per feature. Their ability to handle complex relationships while maintaining high predictive accuracy makes it a valuable tool for both research and clinical applications. By combining the flexibility of non-parametric methods with the simplicity of categorized predictors, biniLasso variants bridge the gap between statistical rigour and practical usability, paving the way for more effective prognostic modelling in high-dimensional settings.

\begin{acks}
The results shown here are in whole or part based upon data generated by the TCGA Research Network: \url{https://www.cancer.gov/tcga}.
\end{acks}

\bibliographystyle{SageH}
\bibliography{refs}

\end{document}